\documentclass[sigconf, review=false, anonymous=false]{acmart}
\usepackage{stmaryrd}
\usepackage{multirow}
\usepackage{amsmath,amsfonts}
\usepackage{graphicx}
\usepackage{enumitem}
\usepackage{textcomp}
\usepackage{./iris}
\usepackage[disable]{todonotes}
\usepackage{natbib}

\DeclareMathOperator*{\astm}{\ast}

\definecolor{obj}{HTML}{984321}
\colorlet{asm}{teal}

\newcommand{\Sail}{\textsc{Sail}}
\newcommand{\muSail}{\textmu\textsc{Sail}}
\newcommand{\Katamaran}{\textsc{Katamaran}}
\newcommand{\Coq}{\textsc{Coq}}
\newcommand{\Iris}{\textsc{Iris}}
\newcommand{\MinimalCaps}{\textsc{MinimalCaps}}

\newcommand{\obj}[1]{{\color{obj}{\mathit{#1}}}}
\newcommand{\instr}[1]{{\color{asm}{\mathrm{#1}}}}
\newcommand{\linenumber}[1]{{\text{\footnotesize {#1}}}}

\newcommand*{\Sref}[1]{\hyperref[#1]{\S\ref*{#1}}}
\newcommand*{\secref}[1]{\hyperref[#1]{Section~\ref*{#1}}}
\newcommand*{\lemref}[1]{\hyperref[#1]{Lemma~\ref*{#1}}}
\newcommand*{\thmref}[1]{\hyperref[#1]{Theorem~\ref*{#1}}}
\newcommand{\corref}[1]{\hyperref[#1]{Cor.~\ref*{#1}}}
\newcommand*{\defref}[1]{\hyperref[#1]{Definition~\ref*{#1}}}
\newcommand*{\egref}[1]{\hyperref[#1]{Example~\ref*{#1}}}
\newcommand*{\appendixref}[1]{\hyperref[#1]{Appendix~\ref*{#1}}}
\newcommand*{\figref}[1]{\hyperref[#1]{Figure~\ref*{#1}}}
\newcommand*{\tabref}[1]{\hyperref[#1]{Table~\ref*{#1}}}

\newcommand{\ie}{\emph{i.e.,} }

\newcommand{\eg}{\emph{e.g.,} }

\newcommand{\wrt}{w.r.t.~}

\newcommand{\braced}[1]{\left\{ #1 \right\}}
\newcommand{\mincapsmachinv}{(\exists c \ldotp (pc \mapsto c) \ast \mathcal{V}(c)) \ast \Sep_{r \in \mathrm{GPR}}\left(\exists w\ldotp r \mapsto w \ast \mathcal{V}(w)\right)}
\newcommand{\multlinedmincapsmachinv}{
\begin{multlined}
\textstyle (\exists c \ldotp (pc \mapsto c) \ast \mathcal{V}(c)) \ast \\
\textstyle \Sep_{r \in \mathrm{GPR}}\left(\exists w\ldotp r \mapsto w \ast \mathcal{V}(w)\right)
\end{multlined}}

\copyrightyear{2023}
\acmYear{2023}
\setcopyright{acmlicensed}\acmConference[CCS '23]{Proceedings of the 2023 ACM SIGSAC Conference on Computer and Communications Security}{November 26--30, 2023}{Copenhagen, Denmark}
\acmBooktitle{Proceedings of the 2023 ACM SIGSAC Conference on Computer and Communications Security (CCS '23), November 26--30, 2023, Copenhagen, Denmark}
\acmPrice{15.00}
\acmDOI{10.1145/3576915.3616602}
\acmISBN{979-8-4007-0050-7/23/11}

\begin{document}

\title{Formalizing, Verifying and Applying ISA Security Guarantees as Universal Contracts}
% copy the following lines to add more authors
% \and
% {\rm Name}\\
% Name Institution
\date{}

\author{Sander Huyghebaert}
\orcid{0000-0002-2878-7429}             %% \orcid is optional
\affiliation{
  % \position{Position2a}
  % \department{Department2a}             %% \department is recommended
  \institution{Vrije Universiteit Brussel}           %% \institution is required
  % \streetaddress{Street2a Address2a}
  % \city{City2a}
  % \state{State2a}
  % \postcode{Post-Code2a}
  \country{}                   %% \country is recommended
}
\affiliation{
  % \position{Position2b}
  % \department{Department2b}             %% \department is recommended
  \institution{KU~Leuven}           %% \institution is required
  % \streetaddress{Street3b Address2b}
  % \city{City2b}
  % \state{State2b}
  % \postcode{Post-Code2b}
  \country{Belgium}                   %% \country is recommended
}
\email{sander.huyghebaert@vub.be}         %% \email is recommended

\author{Steven Keuchel}
\orcid{0000-0001-6411-438X}
\affiliation{
  % \position{Position2a}
  % \department{Department2a}             %% \department is recommended
  \institution{Vrije Universiteit Brussel}           %% \institution is required
  % \streetaddress{Street2a Address2a}
  % \city{City2a}
  % \state{State2a}
  % \postcode{Post-Code2a}
  \country{Belgium}                   %% \country is recommended
}
\email{steven.keuchel@vub.be}          %% \email is recommended

\author{Coen De Roover}
\orcid{0000-0002-1710-1268}             %% \orcid is optional
\affiliation{
  % \position{Position2a}
  % \department{Department2a}             %% \department is recommended
  \institution{Vrije Universiteit Brussel}           %% \institution is required
  % \streetaddress{Street2a Address2a}
  % \city{City2a}
  % \state{State2a}
  % \postcode{Post-Code2a}
  \country{Belgium}                   %% \country is recommended
}
\email{coen.de.roover@vub.be}         %% \email is recommended

\author{Dominique Devriese}
\orcid{0000-0002-3862-6856}
\affiliation{
  % \position{Position2b}
  % \department{Department2b}             %% \department is recommended
  \institution{KU~Leuven}           %% \institution is required
  % \streetaddress{Street3b Address2b}
  % \city{City2b}
  % \state{State2b}
  % \postcode{Post-Code2b}
  \country{Belgium}                   %% \country is recommended
}
\email{dominique.devriese@kuleuven.be}         %% \email is recommendedh.

% Your abstract text goes here. Just a few facts. Whet our appetites.
% Not more than 200 words, if possible, and preferably closer to 150.
\begin{abstract}
  Progress has recently been made on specifying instruction set architectures (ISAs) in executable formalisms rather than through prose.
  However, to date, those formal specifications are limited to the functional aspects of the ISA and do not cover its security guarantees.
  We present a novel, general method for formally specifying an ISA's security
  guarantees to (1) balance the needs of ISA implementations (hardware) and clients (software), (2) can be semi-automatically verified to hold for the ISA operational semantics, producing a high-assurance mechanically-verifiable proof, and (3) support informal and formal reasoning about security-critical software in the presence of adversarial code.
  Our method leverages universal contracts: software contracts that express bounds on the authority of arbitrary untrusted code.
  Universal contracts can be kept agnostic of software abstractions, and strike the right balance between requiring sufficient detail for reasoning about software and preserving implementation freedom of ISA designers and CPU implementers.
  We semi-automatically verify universal contracts against \Sail{} implementations of ISA semantics using our \Katamaran{} tool; a semi-automatic separation logic verifier for \Sail{} which produces machine-checked proofs for successfully verified contracts.
  We demonstrate the generality of our method by applying it to two ISAs that offer very different security primitives: (1) \MinimalCaps{}: a custom-built capability machine ISA and (2) a (somewhat simplified) version of RISC-V with PMP.
  We verify a femtokernel using the security guarantee we have formalized for RISC-V with PMP.
\end{abstract}

\begin{CCSXML}
<ccs2012>
   <concept>
       <concept_id>10011007.10011074.10011099.10011692</concept_id>
       <concept_desc>Software and its engineering~Formal software verification</concept_desc>
       <concept_significance>500</concept_significance>
       </concept>
   <concept>
       <concept_id>10002978.10002986.10002989</concept_id>
       <concept_desc>Security and privacy~Formal security models</concept_desc>
       <concept_significance>500</concept_significance>
       </concept>
   <concept>
       <concept_id>10002978.10002986.10002990</concept_id>
       <concept_desc>Security and privacy~Logic and verification</concept_desc>
       <concept_significance>500</concept_significance>
       </concept>
   <concept>
       <concept_id>10003752.10010124.10010138.10010142</concept_id>
       <concept_desc>Theory of computation~Program verification</concept_desc>
       <concept_significance>500</concept_significance>
       </concept>
   <concept>
       <concept_id>10003752.10010124.10010138.10010140</concept_id>
       <concept_desc>Theory of computation~Program specifications</concept_desc>
       <concept_significance>300</concept_significance>
       </concept>
 </ccs2012>
\end{CCSXML}

\ccsdesc[500]{Software and its engineering~Formal software verification}
\ccsdesc[500]{Security and privacy~Formal security models}
\ccsdesc[500]{Security and privacy~Logic and verification}
\ccsdesc[500]{Theory of computation~Program verification}
\ccsdesc[300]{Theory of computation~Program specifications}

\keywords{universal contracts, ISA security, semi-automatic verification, capability safety, RISC-V, RISC-V PMP}

\maketitle

\section{Introduction}
\label{sec:intro}
An instruction set architecture (ISA) is a contract between software and hardware designers, defining the syntax, semantics, and properties of machine code.
Architecture manuals have traditionally specified the ISA informally through prose.
Such ISA specifications can be imprecise, omit details, and offer no means to test or verify advertised guarantees, which is particularly important for the ISA's security features.
In support of disambiguation, testability, experimentation, and formal study, a recent trend is to instead use formal and executable ISA specifications \citep{armstrong2019isasemantics,bourgeat_multipurpose_2021,flur_modelling_2016,fox_trustworthy_2010,goel_engineering_2017,reid_who_2017,dasgupta_complete_2019}.

For instance, the \Sail{} programming language~\citep{armstrong2019isasemantics} was designed specifically for specifying ISAs.
It is accompanied by a tool that can produce emulators, documentation, and proof assistant definitions from a \Sail{} specification.
\Sail{} has been adopted by the RISC-V Foundation for the official formal specification of RISC-V, an open ISA based on established reduced instruction set computing (RISC) principles \citep{asanovic2014instruction}, and is used for the development of the CHERI extensions~\citep{watson2020capability}.
Furthermore, mature \Sail{} specifications for Armv8a (mechanically translated from authoritative definitions) and RISC-V are available.
Such formal specifications are necessary for formally verifying hardware (processors) and software (compilers, programs written in assembly).

In addition to defining the semantics of instructions, ISA specifications also make statements about the guarantees they uphold.
For example, ISAs offering virtual memory typically guarantee that user-mode code can only access memory that is reachable through the page tables.
Importantly, such guarantees are not just \emph{descriptive} statements
that happen to hold for the current version of the ISA, but \emph{prescriptive} statements which are part of the ISA contract; they must continue to hold for extensions, future versions, and implementations of the ISA.
Similar to the ISA's functional specification, formalizing its security guarantees (rather than just informally specifying them in prose) is vital to support reasoning about security-critical code and validating ISA extensions.

Formalizing ISA security guarantees requires balancing requirements of various stakeholders.
On the one hand, ISA designers and CPU manufacturers require specifications that are abstract and agnostic of software abstractions.
They need to be able to easily validate ISAs and their extensions or updates against the specifications, with maximum assurance.
On the other hand, authors of low-level software need specifications that are sufficiently precise for reasoning about the security properties of code.
They should be able to combine ISA security guarantees, which restrict the authority of untrusted code, with manual reasoning about security-critical, trusted code to obtain full-system security guarantees.

The main contribution of this paper is a general and tool-supported method for formalizing ISA security guarantees, resulting in specifications that are sufficiently abstract to facilitate validating extensions and updates of the ISA, but still sufficiently precise for reasoning about code.
The method is based on so-called \emph{universal contracts} (UCs),
which start from the observation that the ultimate goal of security primitives is to reason about trusted code interacting with untrusted code.
Essentially, the idea is to work in a program logic for assembly code and formulate ISA security guarantees as a universal contract: one that does not just apply to some specific code (as usual in software verification) but a contract that applies to arbitrary ---including untrusted--- code.
This universal contract expresses the restrictions that the ISA enforces on untrusted programs.
When an ISA specification includes a formalization of security guarantees as a UC, it can be used to reason formally about security-critical software.
Manually verified contracts for trusted code can be combined with the universal contract for untrusted code, in order to prove properties about the combined program.

It is essential to verify that the ISA functional specification and security specification are consistent.
In fact, whenever the ISA is extended with additional instructions or behavior,
it is important that these changes do not break the specified security guarantees, to avoid breaking security of software that relies on them.
This means continuously verifying during the evolution of the ISA, that its semantics actually uphold the security properties expressed by the universal contract.

Using universal contracts for generally specifying ISA security guarantees requires (1) a program logic that is sufficiently expressive to express guarantees of a broad range of ISAs, and (2) a verification methodology that can verify the contracts in a semi-automated way against authoritative, engineer-friendly ISA definitions, which are typically implemented as definitional interpreters \citep[e.g., ][]{armstrong2019isasemantics,bourgeat_multipurpose_2021}.
Universal contracts for ISAs have already been used for capturing the capability safety property of capability machine ISAs \citep{georges2021efficient,skorstengaard2018reasoning,van_strydonck_proving_2022}, and integrity and confidentiality properties of Armv7 \citep{dam2013formal, khakpour2013machine}, but neither approach reconciles these two essential requirements.
In the capability machine setting, the universal contracts have a very specific shape: a logical relation is used to define the authority represented by a capability and the universal contract is presented as the fundamental property of the logical relation.
Formulating this logical relation requires a logic that is sufficiently expressive, and as a result, these universal contracts have so far only been proven for theoretical ISAs in a custom-defined but non-authoritative small-step operational semantics, with relatively limited automation.
In the Armv7 setting, the universal contracts are defined using a program logic that is simpler but supports more automation, for a real, authoritative ISA.
However, the more basic program logic used there does not support the reasoning currently done for capability machines and similar work for capability machines has only been able to establish weaker properties \citep{bauereiss2022morello,nienhuis2020rigorous}.
In this paper, we are the first to reconcile the two requirements: logic expressiveness and verification automation for authoritative, engineer-friendly interpreter-style ISA definitions.
More generally, we are the first to propose a general approach to formalizing, verifying and applying the guarantees of more general ISA security primitives.

To achieve this, we use a program logic that allows concise, compositional but expressive proofs and we provide a powerful verification tool called \Katamaran{} to support this: a semi-automatic separation logic verifier for \Sail{} that automates most boilerplate reasoning in the proofs, and allows focusing on the more interesting parts.
\Katamaran{} symbolically executes code in \muSail{} (a core calculus of \Sail{}) to verify general program logic contracts and includes a generic solver for pruning unreachable branches and discharging pure \emph{verification conditions} (VCs).
It is implemented in the \Coq{} proof assistant and comes with a full soundness proof following a general approach \citep{Keuchel2022VerifiedSymbolicExecution}.
This \Coq{} soundness proof provides very high assurance directly against the \muSail{} operational
semantics.
Additionally, taking inspiration from related work \citep{islaris}, \Katamaran{} doubles as a verification tool for the ISA's assembly language, essentially by treating a contract for an assembly program $P$ as a contract for the ISA semantics under the assumption that it is executing program $P$, following related work \citep{islaris}.
Compared to other foundational verifiers like Diaframe \citep{diaframe}, Lithium \citep{refinedc}, MoSeL \citep{mosel} or Bedrock \citep{bedrock}, \Katamaran{} is in a sense a \emph{verified} verification tool rather than a \emph{verifying} one, meaning that it is implemented in Gallina and does not rely on \Coq{} and meta-programming tactics to manage and manipulate intermediate assumptions and VCs.
This approach has benefits for performance and allows extracting the verifier from Gallina to OCaml.
Only a few other tools in the literature take the same approach \citep{verismall,fvf} and they are significantly less practical and mature than \Katamaran{}.

We demonstrate our approach by formalizing the intended security properties for two quite different security primitives: capability safety of a minimalistic capability machine, and memory protection for a version of RISC-V with the Physical Memory Protection (PMP) extension and synchronous interrupts (\ie exceptions).
We prove that the universal contracts hold for the \Sail{}-implemented semantics, the official formal semantics in the case of RISC-V, using \Katamaran{} in a semi-automated approach that improves over the more manual efforts employed so far.
For now, we manually translate the \Sail{} semantics to our internal core calculus \muSail{} and we somewhat simplify the ISA, but the simplifications are minor.
In particular, we fully support bounded integers and byte-addressed memory. The remaining restrictions include assumptions on the amount of PMP configuration registers, the allowed modes for PMP configurations and similar.
The verification tool itself is fully verified, so we obtain high-assurance guarantees in terms of the \muSail{} semantics.

Note that, for now, we only formalize ISA security guarantees about integrity through direct channels, as a first step towards broader guarantees.
In that sense, our work is closely related to recent work on validating security guarantees of capability machine ISAs \citep{nienhuis2020rigorous,bauereiss2022morello} (see Section~\ref{sec:related-work} for a comparison).
Thus, we solve a different problem than recent proposals to make ISAs explicit about side-channel leakage  \citep{ge_time_2019,guarnieri2021contracts}, for which no guarantees are offered by current ISA specifications, formal or informal.

To summarize, the contributions of this paper are:
\begin{itemize}[leftmargin=*,topsep=3pt]
  \item A novel, general method based on universal contracts for formalizing security guarantees of ISAs \wrt the operational semantics of the specification language.
  \item \Katamaran{}: a new semi-automatic tool for verifying separation logic contracts on code in \muSail{} (a new core language for \Sail{}).
        \Katamaran{} supports user-defined abstract predicates, lemma invocations, and heuristics and includes an automatic solver for pure VCs.
        It is implemented and verified in \Coq{}, based on a general approach described elsewhere \citep{Keuchel2022VerifiedSymbolicExecution}.
        Successful verifications produce machine-checked proofs in an \Iris{}-based program logic that is itself proven sound against \muSail{}'s operational semantics.
  \item A demonstration of the method for two case studies: (1) a minimal capability machine that is a subset of CHERI-RISC-V \citep{watson2020capability} and (2) the official formal \Sail{} semantics of RISC-V with the PMP extension, with minor simplifications.
  \item An evaluation of the required effort to validate a UC security guarantee against the operational semantics of an ISA, based on statistics about our two case studies.
        We measure the effort required to validate the addition of an extra instruction.
        To assess the effectiveness of \Katamaran{}'s (semi-)automation, we compare the \MinimalCaps{} verification against a related but more manual proof in Cerise~\citep{georges2021efficient}.
  \item An end-to-end verification demonstrating the usefulness of contracts resulting from our method for reasoning about security-critical code.
        For this purpose we verify an example RISC-V program (called the Femtokernel) that relies on the PMP security guarantees for securing its internal state.
        The verification relies solely on the RISC-V PMP universal contract to reason about the invocation of untrusted code.
\end{itemize}

The remainder of the paper is structured as follows: \secref{sec:background} explains the security primitives used in our case studies.
In \secref{sec:universal-contracts} we introduce universal contracts by formalizing universal contracts for two ISAs.
In \secref{sec:katamaran} we discuss our new semi-automatic logic verifier, \Katamaran{}.
We outline and evaluate the verification effort of our universal contracts in \secref{sec:verif-univ-contr}.
In \secref{sec:femtokernel} we demonstrate the verification of a femtokernel relying on our RISC-V PMP universal contract.
Finally, we discuss related work in \secref{sec:related-work} and conclude in \secref{sec:conclusion}.

\section{Background}
\label{sec:background}
In this section we give a brief introduction to separation logic and cover the security primitives we use in our case studies: capabilities and physical memory protection.

\subsection{Separation Logic}
Program logics are formal frameworks for proving properties about programs.
Hoare logic uses Hoare triples $\{~\mathit{PRE}~\}~ \mathit{P}~ \{~\mathit{POST}~\}$ to express a contract for program P.
The logical predicate $\mathit{PRE}$ expresses a precondition on the initial state of the system (e.g., local variables, heap memory, etc.).
If this precondition is satisfied, then program $\mathit{P}$ is guaranteed to run correctly (e.g., does not get stuck or crash) and if it terminates, the final machine state will satisfy postcondition POST.
If program P returns a result, the postcondition can be written as $\{r\ldotp \mathit{POST}\}$ to express guarantees about the result value $r$, e.g. $\{~\mathit{True}~\}~ \mathrm{return}~42~ \{r\ldotp \mathit{isEven}(r)~\}$.
Separation logic is similar to Hoare logic, but pre- and postconditions are not expressed in standard propositional logic, but in a logic where assertions may also express exclusive ownership of (or authority over) shared mutable state of the system, such as the heap.
This enables more modular reasoning, i.e.\ contracts for program components can more easily be combined.
In separation logic, the points-to predicate $l \mapsto v$ represents ownership of a resource $l$ and knowledge of its current value $v$.
In this paper we will use the points-to predicate for memory locations and registers.
Assertions like $l \mapsto v$ can be combined by separating conjunction, denoted as $P \ast Q$, expressing that $P$ and $Q$ are both true and that they claim ownership of disjoint subsets of the shared state.
The separating implication operator $P \wand Q$ (affectionately referred to as the \emph{magic wand}) requires that $Q$ holds when authority for the premise $P$ is presented.
The separation logic we use in Katamaran is based on \Iris{} \citep{jung2018iris}, which makes the logic particularly expressive.
For example, predicates can be defined recursively (under the restriction of guardedness), they can express authoritative knowledge over advanced forms of so-called ghost state and invariants shared between multiple threads.
Also, a contract $\{~\mathit{PRE}~\}~ \mathit{P}~ \{~\mathit{POST}~\}$ additionally expresses that invariants remain true at every atomic step of execution.

\subsection{Capability Machines}
Capability machines are a special type of processors that offer capabilities.
CHERI is a recent family of capability machine ISA extensions, and includes the Morello ARM extension which is being evaluated in realistic settings by a consortium involving academia and industry \citep{watson_cheri:_2015,watson2020capability}.
Conceptually, capabilities are tokens that carry authority to access memory or an object.
On CHERI, a memory capability extends a traditional pointer with more information such as, amongst others, the bounds and permissions.

\begin{figure}[t]
  \centerline{\includegraphics[width=.7\columnwidth]{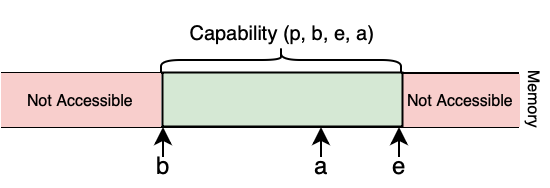}}
  \caption{Concept of a capability}
  \label{capability}
\end{figure}

Capabilities can be represented as a quadruple, $(p, b, e, a)$, consisting of the permission $p$ of the capability, the begin and end addresses $b$ and $e$, and a cursor $a$.
Permissions on a capability machine can include: the null permission $O$, the read permission $R$, and the read and write permission $RW$.
\figref{capability} illustrates a capability's range of authority $[b, e]$ and cursor $a$, pointing to a current memory location.
A special case is the permission $E$, which models enter capabilities\footnote{Also known as sentry capabilities in the context of CHERI \citep{watson2020capability}} \citep{carter_hardware_1994}.
A capability with this permission cannot be used to access memory but can only be jumped to, in which case its permission will change to $R$.
When given to untrusted code, enter capabilities represent a form of encapsulated closures: they can be invoked, but the caller cannot access their private data and capabilities.
As such, they set up a security boundary and can represent a form of software-defined authority and as such they constitute what is generally called an object capability.

The first case studied in this paper is a custom-built capability machine we call \MinimalCaps{}, which supports memory and object capabilities.
It contains a subset of instructions from CHERI-RISC-V \citep{watson2020capability}, including branching, jumping, and arithmetic instructions.
The instructions are a superset of what is supported in Cerise \citep{georges2021efficient,van_strydonck_proving_2022,georges_cap_2021}.
A word on the machine is either an integer or a capability and these can be stored in general-purpose registers (GPRs) and in memory.

\subsection{RISC-V PMP}
The Physical Memory Protection (PMP) extension of RISC-V allows to restrict access to physical address regions \citep{riscv2022spec}.
RISC-V defines three privilege levels: User, Supervisor and Machine (the first two are optional).
PMP allows configuring a memory access policy on 16 or 64 contiguous regions of memory by setting special registers, which are only accessible from the most privileged protection level (machine mode).
PMP has been used to implement a trusted execution environment called Keystone \citep{lee2020keystone}.

\begin{figure}[b]
  \centerline{\includegraphics[width=1\columnwidth]{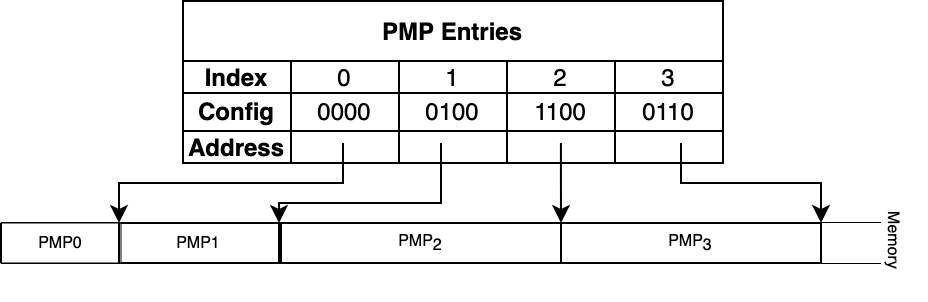}}
  \caption{An example RISC-V PMP policy in Top-of-Range mode (TOR).}
  \label{background:riscv-pmp}
\end{figure}

We illustrate RISC-V PMP policy configuration in \figref{background:riscv-pmp}, where we limit ourselves to four PMP entries.
PMP memory regions are specified by a single address register, which is interpreted according to one of several address-matching modes, but for brevity, we restrict ourselves to Top of Range (TOR).
In TOR mode, the address register of a PMP entry forms the top of the range and the preceding address register (or 0) forms the bottom of the range.
In other words, for PMP entry $i$, the range of the entry is defined as $[\mathit{pmpaddr_{i-1}},\mathit{pmpaddr_i})$, with $\mathit{pmpaddr_{-1}}$ equal to $0$.

In addition to the address, a PMP entry specifies a configuration, which for our purposes consists of 4 bits $LRWX$, where $L$ defines whether the PMP entry is locked and $RWX$ stands for Read, Write and Execute respectively.
The policy in \figref{background:riscv-pmp} grants read-only access to User and Supervisor mode (U- and S-mode) in PMP entry 1 and read-write access in entry 3.
PMP entry 2 is locked, indicating that its \emph{read-only} permission applies to M-mode (machine mode) as well.
Entry 0 grants no permissions and is not locked, so only M-mode can access this range of memory.

We now give a broader explanation of the policy enforced by PMP entries.
By default, M-mode has full permissions over memory while U-mode and S-mode have no permissions.
Non-locked PMP entries grant permissions to U-mode and S-mode.
A locked PMP entry revokes permissions in all modes including M-mode.
Such an entry can only be modified by resetting the system, \ie one cannot write to the associated configuration and address register (and in the case of TOR, the preceding address register).

The PMP check algorithm statically prioritizes the lowest-numbered PMP entries.
For a PMP entry to match an address, all bytes (in the case of multi-byte memory accesses) must match the PMP entry address range.
When a PMP entry matches an address, the LRWX bits determine whether the access succeeds or fails, otherwise the access will succeed in M-mode but fail in other modes.

In our case study we focus on RV32I, the 32-bit base integer instruction set, with the PMP extension.
The case itself is a manual translation from the \Sail{} code to \muSail{}, with some additional simplifications:
only two (rather than 16 or 64) PMP configuration entries in top-of-range (TOR) mode are supported, there is no virtual memory, and we only support M-mode and U-mode.

\subsection{ISA specifications in Sail}
When using \Sail{} an ISA semantics is defined through a definitional interpreter for the ISA's assembly language.
This includes the \emph{Fetch-Decode-Execute cycle} of the ISA.
In the remainder of this paper, we will use the function name \emph{fdeCycle} to refer to the \Sail{} function implementing this cycle, although it may be named differently in a practical \Sail{} specification.
Furthermore, \Sail{} specifications model memory as part of the global state, as well the ISA's registers.
In \figref{fig:sail:fde} we sketch a typical implementation of \emph{fdeCycle}: it indefinitely recurses on invoking the $\mathit{fdeStep}$ function, which fetches, decodes and executes the current instruction.

\begin{figure}[t]
  \begin{align*}
     & \obj{function}~ \mathit{fdeCycle()} = \{ \qquad\qquad\quad \obj{function}~ \mathit{fdeStep()} =                                   \\
     & \quad \mathit{fdeStep()};                \qquad\qquad\qquad\qquad\qquad\quad \obj{let}~ \mathit{w} = \mathit{fetch(PC)}~ \obj{in} \\
     & \quad \mathit{fdeCycle()}                \qquad\qquad\qquad\qquad\qquad\quad \obj{let}~ \mathit{i} = \mathit{decode(w)}~ \obj{in} \\
     & \}                                       \qquad\qquad\qquad\qquad\qquad\qquad\qquad\quad\ \ \mathit{execute(i)}
  \end{align*}
  \caption{$\mathit{fdeCycle}$ and $\mathit{fdeStep}$ definitions as found in \Sail{} specifications.}
  \label{fig:sail:fde}
\end{figure}

To give a better intuition for the $\mathit{execute}$ function, we show the specification of the $\mathit{store}$ instruction in \figref{fig:sail:store}.
\Sail{} models typically define $\mathit{execute}$ as a \emph{scattered definition} that pattern matches on its first argument, the instruction to be executed.
\figref{fig:sail:store} shows such a \emph{clause} for $\mathit{store}$.
The $\mathit{store}$ instruction carries a source register containing the value we want to write into memory, a base register containing the capability to write to memory, and an immediate offset.
The clause performs the write to memory by first reading the capability from register $\mathit{rb}$ and deriving a capability $\mathit{c}$ from $\mathit{bc}$ with its cursor incremented by the immediate.
We assert that the capability has the write permission and then continue by reading the contents of $\mathit{rs}$ and writing it to memory.
Finally, the program counter gets updated.

\begin{figure}[t]
  \begin{align*}
     & \obj{function~clause}~ \mathit{execute(store(rs, rb, immediate))} = \{                                             \\
     & \quad \obj{let}~ \mathit{bc} \mathrel{=} \mathit{read\_reg\_cap(rb)};                                              \\
     & \quad \obj{let}~ \mathit{c} \mathrel{=} \{ \mathit{bc}~ \obj{with}~ \mathit{cap\_cursor} = \mathit{bc.cap\_cursor} \\
     & \quad\quad\quad\quad+ \mathit{sail\_sign\_extend(immediate, integer\_size)} \};                                    \\
     & \quad \obj{assert}(\mathit{writeAllowed(c.cap\_permission)});                                                      \\
     & \quad \obj{let}~ \mathit{w} \mathrel{=} \mathit{read\_reg(rs)};                                                    \\
     & \quad \mathit{write\_mem(c, w)};                                                                                   \\
     & \quad \mathit{update\_pc()};                                                                                       \\
     & \}
  \end{align*}
  \caption{\Sail{} specification of the store instruction of our capability machine (simplified).}
  \label{fig:sail:store}
\end{figure}

\section{Security Guarantees as Universal Contracts}
\label{sec:universal-contracts}

In this section, we explain our novel, general method based on universal contracts in more detail by considering the examples of our capability machine, \MinimalCaps{} and RISC-V PMP.

In our approach, UCs are formulated using separation logic, as introduced in Section~\ref{sec:background}.
Thanks to separation logic, the UCs only have to consider the part of the state an adversary has access to, and can be expected to scale to concurrency which we want to take advantage of in the future.
A contract $\{\mathit{Pre}\}~ \emph{code}~ \{r\ldotp \mathit{Post}\}$ usually applies only to specific $\mathit{code}$.
What we call a universal contract, is one that applies to any choice of \emph{code}, rather than only specific ones.
Such a contract specifies guarantees that are enforced by the semantics for arbitrary or untrusted code.

While universal contracts apply to arbitrary assembly code, they take a slightly different form in our setting.
The arbitrary programs that our contract applies to take the form of arbitrary instructions encoded in memory.
Our universal contract is thus defined as a Hoare triple over the \emph{fdeCycle}.

\subsection{Capability Safety for \MinimalCaps{}}
\label{sec:capability-safety}

The security guarantee offered by a capability machine is capability safety, a property which expresses bounds on the authority of arbitrary untrusted code.
More intuitively, capability safety ensures that arbitrary code can only access the resources and operations for which it has been granted explicit access.
Syntactic properties like capability monotonicity \citep{maffeis_object_2010}, which expresses that the transitively available authority cannot increase during execution, have been shown insufficient in the presence of object capabilities \citep{devriese2016reasoning}.
Therefore, we use a semantic formulation of capability safety based on logical relations \citep{devriese2016reasoning,swasey2017robust,skorstengaard2018reasoning,georges2021efficient,van_strydonck_proving_2022,georges_cap_2021} and formulate the property as a contract over the \emph{fdeCycle}, following Cerise \citep{georges_cap_2021,georges2021efficient,van_strydonck_proving_2022}.
The contract states that if we start from a configuration of safe values, arbitrary code will not be able to exceed the authority of those values.

\figref{fig:lr} shows the logical relation \emph{$\mathcal{V}$} which defines the authority of words (\ie integers and capabilities).
The logical relation is defined using separation logic \citep{reynolds2002separation}, where the notation $\astm\limits_{a \in [b, e]} P$ indicates that $P$ holds separately for all addresses $a \in [b, e]$.

\begin{figure}[t]
  \begin{gather*}
    \mathcal{V}(w) \left\{
    \begin{array}{lll}
      \mathcal{V}(\emph{z})                                        & = & \mathit{True}\qquad (z \in \mathbb{Z})                                             \\
      \mathcal{V}(\emph{O, \textendash, \textendash, \textendash}) & = & \mathit{True}                                                                      \\
      \mathcal{V}(\emph{R, b, e, \textendash})                     & = & \astm\limits_{a \in [b, e]} \knowInv{}{\exists w, a \mapsto w \ast \mathcal{V}(w)} \\
      \mathcal{V}(\emph{RW, b, e, \textendash})                    & = & \astm\limits_{a \in [b, e]} \knowInv{}{\exists w, a \mapsto w \ast \mathcal{V}(w)} \\
      \mathcal{V}(\emph{E, b, e, a})                               & = & \later \always \mathcal{E}(\emph{R,b,e,a})
    \end{array}
    \right.\\
    \mathcal{E}(w)  =
    \left\{\begin{multlined}
      \left(\mathrm{pc} \gmapsto w \ast \Sep_{r \in \mathrm{GPR}}\left(\exists w\ldotp r \mapsto w \ast
        \mathcal{V}(w) \right)\right) \wand\\
      \wpre{\mathit{fdeCycle()}}{~\mathit{True}~}
    \end{multlined}\right.
  \end{gather*}
  \caption{Logical relations for capability safety}
  \label{fig:lr}
\end{figure}

Authority of a value or capability is defined as separation logic predicates that must hold for safely passing the value or capability to untrusted code.
Memory capabilities are thus safe when the addressable locations $a$ are owned by an invariant.
This invariant must require exactly that the word stored at address $a$ always remains safe.
For simplicity, the definition treats read-only capabilities as read-write.
Note that the definition assumes a form of shared invariants, as available in \Iris{}, indicated by a box.
The authority represented by an enter capability is software-defined and therefore non-trivial to define.
Our definition follows previous work and requires that jumping to the capability with its permission changed to $R$ and the general-purpose registers (GPRs) filled with safe words, will execute correctly and will not break any invariants -- that is, $\wpre{\mathit{fdeCycle()}}{~\mathit{True}~}$.
The standard \Iris{} predicate $\wpre{\mathit{fdeCycle()}}{~\mathit{True}~}$ represents the \emph{weakest precondition} for $\mathit{fdeCycle()}$ to execute correctly without breaking any invariants.
It can be alternatively interpreted as $\exists \emph{Pre}, \emph{Pre} \ast \{\emph{Pre}\}~\emph{code}~\{\emph{Post}\}$: some precondition \emph{Pre} needs to hold that is sufficient to guarantee postcondition $\emph{Post}$ after executing $\emph{code}$ \citep{chargueraud2020}.

First-time readers may ignore \Iris{}'s always modality ($\always$), which requires that the authority does not depend on exclusive ownership of resources, and \Iris{}'s later modality ($\later$), which justifies the cycle in the definition of $\mathcal{V}$.
We refer to prior work for more explanation \citep[e.g.,][]{georges2021efficient}.

\begin{figure}
  \centering
  \begin{math}
    \biggl\{\multlinedmincapsmachinv\biggr\}
    \enskip \mathit{fdeCycle()} \enskip
    \biggl\{~\mathit{True}~\biggr\}
  \end{math}
  \caption{Universal Contract for Capability Safety for \MinimalCaps{}.}
  \label{fig:mincapsuniv}
\end{figure}

The universal contract for \MinimalCaps{} is a contract for the fetch-decode-execute cycle, depicted in \figref{fig:mincapsuniv}.
It asserts that if the machine is executed (\ie \textit{fdeCycle()} is invoked) with authoritized capabilities in \textrm{pc} and general-purpose registers, then it will execute correctly and not break any invariants.
The postcondition \textit{True} is trivial, as before when we used $\wpre{\mathit{fdeCycle()}}{~\mathit{True}~}$.
Indeed, $\mathit{fdeCycle()}$ is an infinite loop, so the postcondition is not very relevant anyway.
Nevertheless, even with an irrelevant postcondition, our contracts still express the preservation of invariants during execution (as we will see in Section~\ref{sec:femtokernel}).

As demonstrated before \citep{van_strydonck_proving_2022,georges2021efficient,georges_cap_2021}, such a UC is agnostic of software abstractions but supports reasoning about untrusted code.
Essentially, one can register integrity properties of trusted code as invariants\footnote{In fact, \Iris{} invariants can also express protocols on private state \citep{jung2018iris}.}, and then use the UC for justifying jumps to untrusted code.
Applying the UC requires proving that authority is available for all words that the untrusted code gets access to, directly (in a register) or indirectly (in memory reachable from register capabilities).
This includes proving that enter capabilities passed to the adversary can be invoked freely but never break established invariants.

\subsection{Memory Integrity for RISC-V PMP}
\label{sec:risc-v-pmp}
Universal contracts can be applied beyond the specific setting of capability machines.
While other work has already employed universal contracts for specific non-capability machines, namely Armv7 \citep{dam2013formal,khakpour2013machine}, we provide evidence that our method and our tool \Katamaran{} are general and can be used to capture security guarantees of different security primitives.
To this end we apply our approach to RISC-V with support for exceptions (synchronous interrupts) and the PMP extension, as explained in \secref{sec:background}.
The universal contract captures the memory integrity guarantee offered by the ISA when invoking untrusted code.

Our model of RISC-V is translated to \muSail{} from the ISA's canonical \Sail{} semantics with some minor simplifications:
only two (rather than 16 or 64) PMP configuration entries in top-of-range (TOR) mode are supported, no virtual memory, and only M-mode and U-mode (no S-mode).

\begin{figure}[t]
  \begin{math}
    \begin{array}{l}
      \left\{
      \begin{array}{l}
        \mathit{Normal}(l,h,\mathit{mpp},\mathit{entries})                                               \\
        \astm \later (\mathit{CSR\ Modified}(l,\mathit{entries}) \wand \wpre{\mathit{fdeCycle()}}{~\mathit{True}~}) \\
        \astm \later (\mathit{Trap}(l,h,\mathit{entries}) \wand \wpre{\mathit{fdeCycle()}}{~\mathit{True}~})        \\
        \astm \later (\mathit{Recover}(l,h,\mathit{mpp},\mathit{entries}) \wand \wpre{\mathit{fdeCycle()}}{~\mathit{True}~})
      \end{array}
      \right\} \\
      \mathit{fdeCycle()} \left\{
      ~\mathit{True}~
      \right\}
    \end{array}
  \end{math}%
  \begin{multline*}
    \mathit{Normal}(l,h,\mathit{mpp},\mathit{entries}) =\\
    \left\{
    \begin{alignedat}{1}
      &(\exists i\ldotp \mathit{pc}             \mapsto i) \astm
      \mathit{cur\_privilege} \mapsto l \astm
      \mathit{mtvec}          \mapsto h \astm\\
      &(\exists c\ldotp \mathit{mcause}         \mapsto c) \astm
      \mathit{mstatus}        \mapsto [ mpp ] \astm \\
      &\mathit{mepc}           \mapsto \mathit{mepc} \astm
      \mathit{PMP\_entries\ entries} \astm \\
      &\Sep_{r \in \mathrm{GPR}}\left(\exists w\ldotp r \mapsto w \right) \astm
      \mathit{PMP\_addr\_access\ entries\ l}
    \end{alignedat} \right.
  \end{multline*}%
  \begin{multline*}
    \mathit{Trap}(l,h,\mathit{entries}) =\\
    \left\{
    \begin{alignedat}{1}
      &\mathit{pc}             \mapsto h \astm
      \mathit{cur\_privilege} \mapsto \mathit{Machine} \astm
      \mathit{mtvec}          \mapsto h \astm\\
      &(\exists c\ldotp \mathit{mcause}         \mapsto c) \astm
      \mathit{mstatus}        \mapsto [ l ] \astm \\
      &(\exists c\ldotp \mathit{mepc}           \mapsto c) \astm
      \mathit{PMP\_entries\ entries} \astm \\
      &\Sep_{r \in \mathrm{GPR}}\left(\exists w\ldotp r \mapsto w \right) \astm
      \mathit{PMP\_addr\_access\ entries\ l}
    \end{alignedat}
    \right.
  \end{multline*}

  \caption{Universal Contract for Memory Integrity for RISC-V with PMP.}
  \label{fig:riscvcontract}
\end{figure}

We define the universal contract for this machine, over the fetch-decode-execute cycle as shown in \figref{fig:riscvcontract}.
In this contract the machine starts from a Normal state, which requires ownership (and knowledge of the current values) of the architectural registers $\mathit{pc}$, $\mathit{cur\_privilege}$, $\mathit{mtvec}$, $\mathit{mcause}$, $\mathit{mstatus}$ and $\mathit{mepc}$, containing respectively the program counter, current privilege level, configured exception handler address, cause of the last interrupt, and the privilege level and program counter before the last interrupt.
Additionally, the state requires ownership of the general-purpose registers and the current PMP configuration entries.
Finally and perhaps most importantly, \textbf{Normal} requires ownership of PMP\_addr\_access entries l, a predicate that represents ownership of memory accessible according to the PMP policy entries at the current privilege l (discussed further below).

Given this authority, the contract states that the ISA will execute correctly, provided that three extra conditions are fulfilled.
All three require that the machine continues executing correctly in a specific situation: (1) when CSRs are modified, (2) when a trap occurs to the exception handler, and (3) when an MRET is used to return to a lower privilege level (Recover).

For brevity, we only show the definition of Trap, arguably the most important case, as the other two cases can only be reached if the original privilege level l was Machine, \ie{} the UC is used to reason about untrusted Machine code.
PMP can be used to encapsulate Machine code (by locking some PMP entries) but it is more typically used for encapsulating lower-privilege code.
Trap requires ownership of the same ISA registers and memory as Normal above.
However, it additionally requires that the program counter is set to the configured exception handler, that cur\_privilege is set to Machine mode, and that mstatus correctly stores l as the previous privilege level.
Under these conditions, the user of the UC needs to prove that the machine will execute correctly.
This reflects the intuition that trusted code can rely on PMP to encapsulate untrusted lower-privilege code, but only if it ensures security of the configured exception handler.

\begin{figure}[b]
  \begin{math}
    \begin{array}{l}
      \mathit{PMP\_addr\_access\ entries\ m} = \\
      \quad \Sep_{a \in addrs}
      ((\exists p, PMP\_access\ a\ entries\ m\ p) \wand \exists w, a \mapsto w)
    \end{array}
  \end{math}

  \caption{PMP\_addr\_access predicate implementation}
  \label{fig:pmpaddraccess}
\end{figure}

A crucial predicate in the universal contract is $\mathit{PMP\_addr\_access}$, which captures the semantics of the PMP check algorithm and is shown in \figref{fig:pmpaddraccess}.
It is defined as a separating conjunction over all addresses of the machine.
The predicate allows obtaining a pointsto predicate for an address $a$, if the PMP policy specifies a permission $p$ (e.g.\ \emph{Read} or \emph{Write}) for it at privilege level l.
Importantly, this means that ownership of other memory locations is not required for using the universal contract, so it cannot be accessed.

\subsection{Verifying and applying a Universal Contract}
Formalizing security guarantees is useful for two purposes.

First, it can be used to verify whether the security guarantees are consistent with the ISA's operational semantics, \eg whether all instructions correctly check the (PMP or capability-based) policy before accessing memory.
This verification is a non-trivial effort that should be repeatable during the evolution of the ISA, so we provide tool support in the form of \Katamaran{}, a semi-automatic verification tool for \Sail{}, discussed in Section~\ref{sec:katamaran}.
We discuss verifying the UCs for \MinimalCaps{} and RISC-V PMP in Section~\ref{sec:verif-univ-contr}.

Secondly, the UC can be used to verify the security of trusted software interacting with untrusted software.
If such software is verified against the UC, its security will hold in any implementation of the ISA that respects the UC.
Such a verification entails verifying trusted code using the ISA operational semantics and invoking the UC when a jump to untrusted code happens.
We demonstrate in Section~\ref{sec:femtokernel} that our UC supports this using the \emph{femtokernel} case study: a minimal RISC-V PMP machine mode kernel that interacts with untrusted user-mode code over a simple system call.
Interestingly, this verification reuses \Katamaran{} as a verification tool for RISC-V assembly, by reusing an idea from previous work \citep{islaris}.

\section{Katamaran}
\label{sec:katamaran}

Verifying that the semantics upholds security properties is a significant
endeavor which involves manual reasoning.
For instance, the \Coq{} formalization of \citet{georges_cap_2021}'s capability safety proof for a simple capability machine with 19 instructions requires about
7k\textsc{loc} of \Coq{} proofs.
Real-world ISAs can of course be much larger. Consequently, scaling
up verification of ISA properties raises important proof engineering challenges.
Furthermore, if the ISA changes (because of minor updates, new features, or for experimentation), the proofs have to be updated as well.
For manual proofs, this can result in a prohibitive amount of work.

In a nutshell, proof automation is mission-critical for the verification effort
to scale in terms of the size and complexity of the specification of
the instruction sets and of the specification of the security guarantee itself,
and for proofs to be robust to changes in the specification.

Proof automation means that uninteresting or repetitive parts of the proof are
dealt with automatically using a tool, library, script etc.
The goal is for a human to steer the automation by providing heuristics, and she should also be able to intervene directly and prove certain cases manually where full automation fails.
In other words, verifying security properties of ISAs should at least be semi-automatic.

To this end, we have developed \Katamaran{}, a new semi-automatic separation logic verifier, implemented and proven sound using Kripke specification monads~\citep{Keuchel2022VerifiedSymbolicExecution}.
\Katamaran{} is developed as a library for the \Coq{} proof assistant, and
works with \muSail{}, a new core calculus for \Sail{} deeply embedded in \Coq{},
offering many of \Sail{}'s features.\footnote{\Sail{}'s existing \Coq{} backend only translates to a shallow embedding.}
For the time being, the translation from \Sail{} to \muSail{} has to be performed manually, but we intend to automate it in the future.

Much like \Sail{}, \muSail{} specifications also leave the definition of memory out of the functional
specification and require a (user-provided) runtime system to define what
constitutes the machine's memory and to provide access to it.
To this end, \Katamaran{} relies on foreign functions -- that is, functions implemented in \Coq{} of which the signature has been declared in \muSail{} so they are callable. Additionally, \muSail{} allows invoking lemmas (sometimes referred to as ghost statements), which instructs the verifier to take a non-trivial proof step that is verified separately.

The security properties are specified by means of separation logic-based
contracts consisting of pre- and postconditions for all functions, including
foreign ones. For this, \Katamaran{} contains its own deeply embedded assertion
language.

\begin{figure}[b]
  \centerline{\includegraphics[width=1.0\columnwidth]{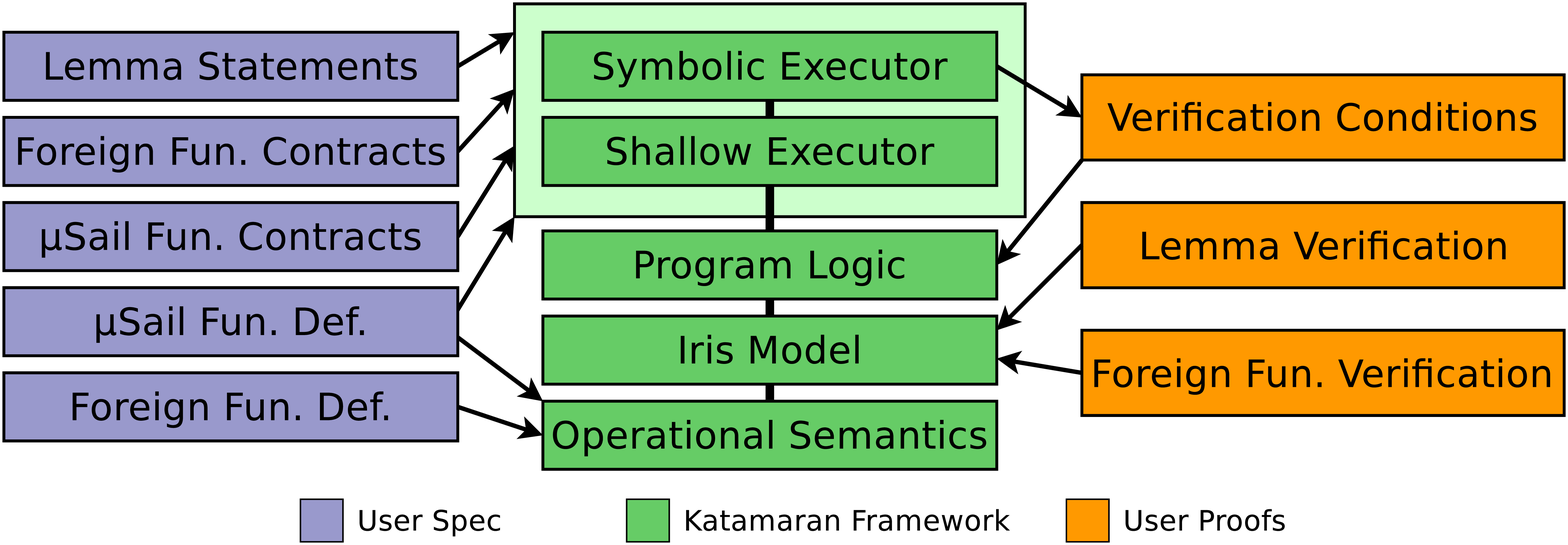}}
  \caption{Structure of \Katamaran{}}
  \label{fig:katamaran}
\end{figure}

Verifying that functions adhere to their contracts is done via
\emph{preconditioned forward static symbolic execution}
\citep{surveyofsymbolicexecution,berdine2005} of the function bodies. During the
execution, \Katamaran{} tries to discharge proof obligations automatically using solvers and leaves residual VCs for the user where this fails.
The library contains a \emph{generic solver} for some of \muSail{}'s background theory, which can be complemented by a \emph{user solver} for user-defined predicates.
To bound the burden, we
require that all spatial proof obligations -- that is, those related to registers and memory, are dealt with during symbolic execution, potentially with the help of the user in
terms of ghost statements and heuristics, and thus only pure proof obligations remain. %\domi{and heuristics?}.
Hence, the produced residual VCs will be in first-order predicate logic, which the user can discharge using \Coq{}'s built-in proof automation.

\Katamaran's symbolic executor is implemented as a monadic interpreter in a specification
monad~\cite{swamy2013verifyingwithdijkstramonad,ahman2017dijkstramonadsforfree,dijkstramonadsforall}.
Such specification monads allow angelic and demonic non-determinism which we use to explore the execution paths of \Sail{} programs.
The specification monad is implemented as a predicate transformer, which combines assertions and assumptions encountered on different execution paths into a verification condition.
The resulting verification conditions are not regular \Coq{} propositions but rather ASTs in a syntactically represented language of propositions, allowing us to simplify them and prune unreachable paths during verification.
Additionally, we use Kripke indexing techniques to track logical variables in scope and path constraints \citep{Keuchel2022VerifiedSymbolicExecution}.
This enforces that path constraints are monotonically increasing along execution paths and hides the plumbing.

A question that arises is whether the generated VCs suffice to verify the function contracts.
The user does not have to take the output of the symbolic executor at face value: \Katamaran{} comes with
a full soundness proof against the \muSail{} operational semantics.
The structure is depicted in \figref{fig:katamaran}.
The contracts of both kinds of functions and the code of the \muSail{} functions are inputs to the symbolic executor from which it produces VCs.
Following the recipe proposed by~\citep{fvf,Keuchel2022VerifiedSymbolicExecution}, we first reduce the soundness of the symbolic executor to a shallow executor: a monadic interpreter with the same structure as the symbolic executor but written in a specification monad over propositions of the meta-logic.
A binary Kripke logical relation~\citep{Keuchel2022VerifiedSymbolicExecution} links the constituents of both interpreters, and allows us to conclude that symbolic VCs imply shallow VCs.
A second soundness proof connects this to an axiomatic program logic: given proofs of the shallow VCs the function bodies are also verifiable in the program logic.

The program logic consists of separation logic contracts \linebreak $\{~\mathit{PRE}~\}~c~\{r\ldotp \mathit{POST}\}$.
We assign meaning to these contracts using the \Iris{} separation logic framework \citep{jung2018iris}.
This requires user-provided proofs that foreign functions adhere to their contracts and that lemmas used in ghost statements are sound.
We kept the axiomatic program logic separate from its instantiation using \Iris{}, and in principle, other logics than \Iris{} can be used.
However, we provide the \Iris{} model as the default choice with full soundness proofs and hooks for the user to extend it.

A last adequacy proof connects the \Iris{} contracts to the operational semantics: every contract that holds semantically implies partial correctness.
This is sufficient for us; we assume it is verified separately that the machine cannot get stuck.

\begin{table}[t]
  \caption{Katamaran lines of code calculated by \texttt{coqwc}.}\label{tab:katamaran}
  \centering
  \begin{tabular}{lll}
    \toprule
    Component                      & Spec  & Proof \\
    \midrule
    \muSail{} syntax and semantics & 1939  & 547   \\
    Symbolic executor              & 3102  & 2396  \\
    Background theory solver       & 786   & 410   \\
    Shallow executor               & 536   & 762   \\
    Program logic                  & 1226  & 1058  \\
    \Iris{} model                  & 596   & 679   \\
    Other                          & 2609  & 1325  \\
    \bottomrule
    Total                          & 10794 & 7177  \\
  \end{tabular}
\end{table}

An overview of the sizes of different parts of \Katamaran{} can be found in \tabref{tab:katamaran}.
Even though we run \Katamaran{} within \Coq{}'s typechecker\footnote{Extracting or natively compiling the code is possible in theory, but currently the overhead outweighs the benefits.}, it is fast enough for interactive experimentation with definitions in our case studies, with immediate verification of the corresponding contracts.
For instance, the longest runs of the symbolic executor are the verification of the femtokernel blocks that we describe in \secref{sec:femtokernel}, where we symbolically execute \muSail{} code to derive a symbolic executor for assembly code.
Both blocks combined take 1.77s to symbolically execute (on an Intel i7-12700) and in total 247 calls of \muSail{} and foreign functions are executed of which 105 calls are executed by interpreting function contracts and 142 are executed by interpreting function bodies.

\section{Verification of Universal Contracts using \Katamaran{}}
\label{sec:verif-univ-contr}

Using \Katamaran{}, we have verified that the two universal contracts from Section~\ref{sec:universal-contracts} are consistent with the operational semantics of the ISAs.
In this section, we explain the two verifications and evaluate the proof effort involved.

\subsection{\MinimalCaps{}}
The verification of capability safety in the literature so far has required significant manual effort \citep{georges2021efficient,skorstengaard2018reasoning,van_strydonck_proving_2022}.
In this section, we demonstrate our semi-automatic approach.

The contract for \textit{fdeCycle()} iterates the following contract for \textit{fdeStep()}, which executes a single FDE cycle.
\begin{gather*}
  \textstyle\braced{\mincapsmachinv \ast \mathit{IH}} \\
  \mathit{fdeStep()} \\
  \textstyle\braced{(\exists c \ldotp (pc \mapsto c) \ast \mathcal{V}(c) \vee \mathcal{E}(c)) \ast \Sep_{r \in \mathrm{GPR}}\left(\exists w\ldotp r \mapsto w \ast \mathcal{V}(w)\right)}
\end{gather*}
This internal contract requires an induction hypothesis IH:
\begin{multline*}
  \mathit{IH} \mathbin{:=} \always \later (\forall c\ldotp pc \mapsto c \ast \mathcal{V}(c) \ast\\
  \Sep_{r \in \mathrm{GPR}}\left(\exists w\ldotp r \mapsto w \ast \mathcal{V}(w)\right) \wand \wpre{\mathit{fdeCycle()}}{~\mathit{True}~})
\end{multline*}
Note how the postcondition allows the pc to contain a safe capability or one that satisfies the expression relation $\mathcal{E}$ above.
The latter is necessary because after invoking an enter capability, the pc may contain a value that would not be safe to hand to an adversary but is nevertheless safe to execute.
We apply the same contract to helper functions used by $\mathit{fdeStep()}$ to execute individual instructions.
Some other helper functions are given more specific contracts.

Consider, for example, the $\mathit{read\_mem}(c)$ function, which reads the word in memory denoted by the cursor of the given capability.
The contract of $\mathit{read\_mem}$ requires authority for capability $(p,b,e,a)$ before executing $\mathit{read\_mem}(c)$ with permission $p$ at least read permission, and guarantees that the capability is still safe afterwards, as well as the word read from memory:
\begin{multline*}
  \braced{\mathcal{V}(p,b,e,a) \ast R \leq_p p} \mathit{read\_mem~(p,b,e,a)}\\ \braced{w \ldotp \mathcal{V}(w) \ast \mathcal{V}(p,b,e,a)}
\end{multline*}

\begin{figure}[t]
  \begin{align*}
     & \textcolor{red}{\{ (\exists c\ldotp pc \mapsto c \ast \mathcal{V}(c)) \Sep_{r \in \mathrm{GPR}}\left(\exists w\ldotp r \mapsto w \ast \mathcal{V}(w)\right)} \}                                     \\
     & \obj{store} (\mathit{rs} : \mathrm{GPR}, \mathit{rb} : \mathrm{GPR}, \mathit{immediate} : \mathrm{int}) : \mathrm{bool} \mathrel{:=}                                                                \\
     & \quad \mathbf{let}~ \mathit{bc} \mathrel{:=} \mathbf{call}~\obj{read\_reg\_cap}~rb~ \mathbf{in}                                                                                                     \\
     & \quad \mathbf{let}~ (\mathit{perm},~ \mathit{beg},~ \mathit{end},~ \mathit{cursor}) \mathrel{:=} \mathit{bc}~ \mathbf{in}                                                                           \\
     & \quad \mathbf{let}~ c \mathrel{:=} (\mathit{perm},~ \mathit{beg},~ \mathit{end},~ \mathit{cursor} + \mathit{immediate})~ \mathbf{in}                                                                \\
     & \quad \mathbf{let}~ p \mathrel{:=} \mathbf{call}~\obj{write\_allowed}~\mathit{perm}~ \mathbf{in}                                                                                                    \\
     & \quad \mathbf{assert}~ p;                                                                                                                                                                           \\
     & \quad \mathbf{let}~ w \mathrel{:=} \mathbf{call}~\obj{read\_reg}~rs~ \mathbf{in}                                                                                                                    \\
     & \quad \mathbf{lemma}~ \obj{subperm\_not\_E}~\mathrm{RW}~\mathit{perm};                                                                                                                              \\
     & \quad \textcolor{red}{\braced{r_0 \mapsto \mathit{bc} \ast \mathcal{V}(\mathit{bc}) \ast r_1 \mapsto w_1 \ast \mathcal{V}(w_1) \ast \mathit{perm} \neq \mathrm{E} \ldots}}                          \\
     & \quad \mathbf{lemma}~ \obj{move\_cursor\ bc}~c;                                                                                                                                                     \\
     & \quad \textcolor{red}{\braced{r_0 \mapsto \mathit{bc} \ast \mathcal{V}(\mathit{bc}) \ast r_1 \mapsto w_1 \ast \mathcal{V}(w_1) \ast \mathit{perm} \neq \mathrm{E} \ast \mathcal{V}(c) \ast \ldots}} \\
     & \quad \mathbf{call}~ \obj{write\_mem}~c~w;                                                                                                                                                          \\
     & \quad \mathbf{call}~ \obj{update\_pc};                                                                                                                                                              \\
     & \textcolor{red}{\{ (\exists c\ldotp pc \mapsto c \ast \mathcal{V}(c)) \ast \Sep_{r \in \mathrm{GPR}}\left(\exists w\ldotp r \mapsto w \ast \mathcal{V}(w)\right) \}}
  \end{align*}
  \caption{Capability safety for the store instruction (slightly simplified).}
  \label{fig:mincaps:store}
\end{figure}

\begin{figure}[b]
  \begin{align*}
    \braced{\mathcal{V}(p,b,e,a) \ast R \leq_p p}           & \,\mathit{read\_mem~(p,b,e,a)} \hookleftarrow                                                   \\
                                                            & \, \braced{w \ldotp \mathcal{V}(w) \ast \mathcal{V}(p,b,e,a)}                                   \\
    \braced{r \mapsto w}                                    & \,\mathit{read\_reg\ r}\,                 \braced{v \ldotp v = w \ast r \mapsto w}              \\
    \braced{r \mapsto w}                                    & \,\mathit{read\_reg\_cap\ r}\,            \braced{c \ldotp c = w \ast r \mapsto w}              \\
    \braced{\mathcal{V}(c) \ast \mathcal{V}(w)}             & \,\mathit{write\_mem\ c\ v}\,             \braced{\mathcal{V}(c)}                               \\
    \braced{pc \mapsto c \ast \mathcal{V}(c)}               & \,\mathit{update\_pc}\,                   \braced{\exists c . pc \mapsto c \ast \mathcal{V}(c)} \\
    \braced{\mathcal{V}(p, b, e, a) \ast p \neq \mathrm{E}} & \,\mathit{move\_cursor\ (p, b, e, a)\ (p, b, e, a')}\hookleftarrow                              \\
                                                            & \qquad\braced{\mathcal{V}(p, b, e, a) \ast \mathcal{V}(p, b, e, a')}                            \\
    \braced{
      \begin{multlined}
        (p = \mathrm{R} \vee p = \mathrm{RW})\\ \ast p \leq_p p'
      \end{multlined}
    }                                                       & \,\mathit{subperm\_not\_E\ p\ p'} \braced{p' \neq \mathrm{E}}
  \end{align*}
  \caption{Contracts for functions and lemmas used in exec\_sd (for registers r, values v and w and capabilities c)}
  \label{fig:contracts}
\end{figure}

To give an idea of how these contracts are verified using \Katamaran{}, \figref{fig:mincaps:store} shows the \muSail{} implementation of \MinimalCaps{}' store instruction, with verification annotations in red (not part of the code itself), and \figref{fig:contracts} displays the contracts for the functions used in the implementation.
This instruction takes 3 arguments: source and target GPRs $\mathit{rs}$ and $\mathit{rb}$ and an integer $\mathit{immediate}$.
The instruction will write the value of $\mathit{rs}$ to $\mathit{cursor} + \mathit{immediate}$, if register $\mathit{rb}$ contains a capability with this $\mathit{cursor}$.

In the function body, a new capability $c$ is derived from $\mathit{bc}$ with the immediate added to the cursor, and this capability is used to write word $w$ to memory.
We use a few lemmas to modify the precondition in order to satisfy the precondition of $\mathit{write\_mem}$, which requires authority for destination capability $c$ and word $w$.
Note that lemmas are proven sound in the \Iris{} model for the case study using \Iris{} Proof Mode.
For simplicity, we assume that $\mathit{rb} = R0, \mathit{rs} = R1$ and ignore the non-relevant parts of the precondition for this discussion.

The $\mathit{move\_cursor}$ lemma produces a $\mathcal{V}$ predicate for $c$ based on the one for $\mathit{bc}$, which differs only in the cursor field.
Remember that the authority of memory capabilities requires that all addresses between $[begin, end]$ are owned and point to words whose authority is available, \ie it does not mention the cursor of the capability.
Because move\_cursor only works for non-$\mathrm{E}$ capabilities, we use another lemma subperm\_not\_E to derive that $\mathit{perm} \neq \mathrm{E}$ from $\mathrm{RW} \leq_p \mathit{perm}$.

The $\mathit{write\_mem}~c~w$ call checks that $c$'s cursor is within bounds but assumes it has the write permission.
If all checks pass, word $w$ will be written to this address.
The checks are critical for capability safety of \MinimalCaps{} and the machine will go into a failed state if they are not satisfied.
The actual write to memory is performed through a foreign function, called $\mathit{wM}$, which takes an address and a word to be written to memory.
$\mathit{wM}$ is provided by the \Sail{} standard library for the \Sail{} specification and in the runtime system for its \muSail{} counterpart.

The $\mathit{update\_pc}$ function is quite simple and utilizes the \linebreak $\mathit{move\_cursor}$ lemma again to generate a $\mathcal{V}$ predicate for the updated pc.

Arriving at the end of the $\mathit{exec\_sd}$ function, we can verify that its contract holds, \ie safety of register values
is preserved when executing this instruction.
Together with the verification of other functions, we derive the contract for \textit{fdeCycle()}.
The contract for the \textit{fdeCycle()} itself is proven manually due to the use of \Iris{}'s later modality and L\"ob induction, which \Katamaran{} does not (yet) support.
Note that in the proof of the contract of \textit{fdeCycle()}, we can use the semi-automatically proven contracts, \ie we don't need to reason about \textit{fdeStep()} manually in the body of \textit{fdeCycle()}.
The \textit{fdeCycle()} contract is a universal contract of the ISA, as it expresses an authority boundary on (untrusted) code.
It allows us to conclude that our \MinimalCaps{} ISA actually satisfies the intended capability safety property.

\subsection{RISC-V PMP}
\label{sec:riscv-verification}

\begin{figure}%[b]
  \footnotesize
  \begin{alignat*}{2}
    \left\{
    \begin{array}{l}
      \mathit{Read} \sqsubseteq t                  \\
      \astm \mathit{cur\_privilege} \mapsto p      \\
      \astm \mathit{PMP\_entries\ entries}         \\
      \astm \mathit{PMP\_access\ a\ entries\ p\ t} \\
      \astm a \mapsto w
    \end{array}
    \right\}
     & \,\mathit{read\_ram\ a}\,     &
    \left\{\begin{array}{l}
             \mathit{cur\_privilege} \mapsto p    \\
             \astm \mathit{PMP\_entries\ entries} \\
             \astm a \mapsto w
           \end{array} \right\} \\
    \left\{
    \begin{array}{l}
      \mathit{Write} \sqsubseteq t                 \\
      \astm \mathit{cur\_privilege} \mapsto p      \\
      \astm \mathit{PMP\_entries\ entries}         \\
      \astm \mathit{PMP\_access\ a\ entries\ p\ t} \\
      \astm \exists w, a \mapsto w
    \end{array}
    \right\}
     & \,\mathit{write\_ram\ a\ v}\, &
    \left\{\begin{array}{l}
             \mathit{cur\_privilege} \mapsto p    \\
             \astm \mathit{PMP\_entries\ entries} \\
             \astm a \mapsto v
           \end{array} \right\}
  \end{alignat*}
  \caption{Contracts for functions interacting directly with memory}
  \label{fig:ramcontracts}
\end{figure}

As for \MinimalCaps{}, we verify that our universal contract holds for the functional specification of RISC-V.
This verification is done assuming contracts for reading from and writing to memory, shown in \figref{fig:ramcontracts}.
The contracts require read or write access, respectively in the form of the $\mathit{PMP\_access}$ predicate encountered above.
We also require that we have ownership of the address that we want to read from or write to, $a \mapsto w$.
The postconditions of these functions return the resources used, updated in the case of $\mathit{write\_ram}$ to point to the newly written value.

\begin{figure*}[t]
  \centerline{\includegraphics[width=2.2\columnwidth]{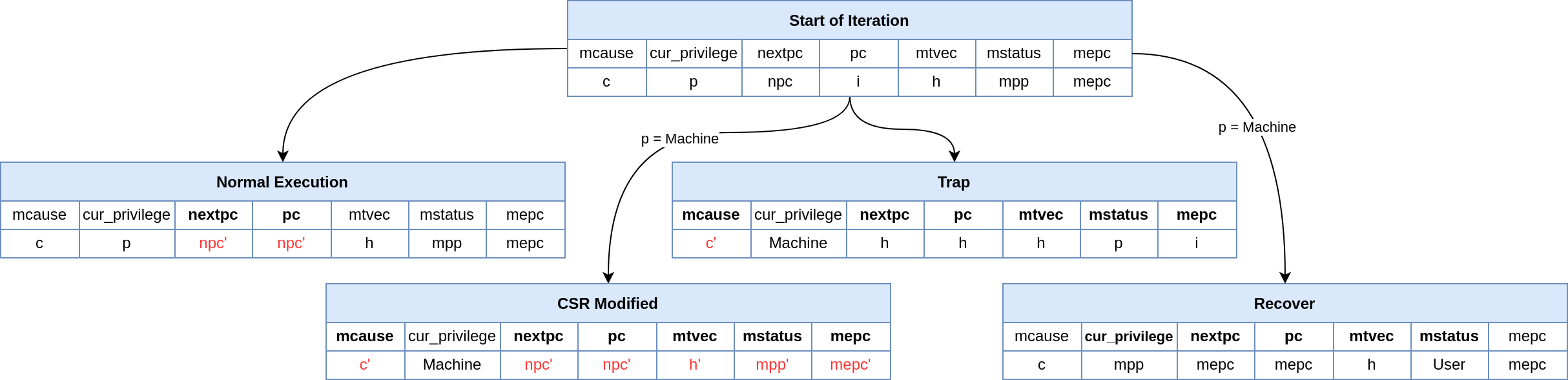}}
  \caption{Contract for taking a step on RISC-V (\ie executing an instruction). New existentially quantified logic variables are shown in red, modified registers are shown in bold.
    Constraints on the Start of Iteration logic variables are indicated on the arrows (we require for CSR Modified and Recover state transitions that we started from a state running
    in Machine mode, \ie $p = \mathit{Machine}$).}
  \label{fig:instrexeccontract}
\end{figure*}

Like for \MinimalCaps{}, the universal contract proof iterates a contract for the single-cycle \textit{fdeStep()}, depicted in \figref{fig:instrexeccontract}.
It specifies that executing an instruction will leave the CPU in one of the states Normal, CSRModified, Trap or Recover, mentioned earlier, with specific values for the ISA registers.
Not shown are the predicates for ownership over the general-purpose registers, \ie $\Sep_{r \in \mathrm{GPR}}\left(\exists w\ldotp r \mapsto w \right)$, and the PMP entries, $\mathit{PMP\_entries}$, and $\mathit{PMP\_addr\_access}$, representing ownership of the PMP-authorized memory.
All these predicates are preserved as-is upon a state change, except $\mathit{PMP\_entries}$ which may be modified in the \textbf{CSR Modified} state.
The CSRModified and Recover states can only be reached when executing in \emph{Machine} mode, \ie $p = \mathit{Machine}$.
Trap transfers into \emph{Machine} mode and Recover returns to the privilege level stored in mpp.

\begin{figure*}[t]
  \footnotesize
  \begin{math}
    \begin{array}{r l l}
      \braced{\mathit{PMP\_entries\ entries}}
       & \,\mathit{open\_PMP\_entries}\,  &
      \left\{
      \begin{array}{l}
        \exists \mathit{cfg_0,\ addr_0,\ cfg_1,\ addr_1,\ } \\
        (\mathit{pmp_0cfg} \mapsto \mathit{cfg_0} \astm
        \,\mathit{pmpaddr_0} \mapsto \mathit{addr_0} \astm  \\
        \,\mathit{pmp_1cfg} \mapsto \mathit{cfg_1} \astm
        \,\mathit{pmpaddr_1} \mapsto \mathit{addr_1} \astm  \\
        \,\mathit{entries} = [(\mathit{cfg_0}, \mathit{addr_0}); (\mathit{cfg_1}, \mathit{addr_1})])
      \end{array}
      \right\}                              \\[1em]
      \left\{
      \begin{array}{l}
        \mathit{PMP\_addr\_access\ entries\ p} \astm \\
        0 \leq \mathit{addr} \leq \mathit{maxAddr} \astm \mathit{PMP\_access\ addr\ entries\ p\ acc}
      \end{array}
      \right\}
       & \,\mathit{extract\_PMP\_ptsto}\, &
      \left\{
      \exists w\ldotp \mathit{addr} \mapsto w \astm (\mathit{addr} \mapsto w \wand \mathit{PMP\_addr\_access\ entries\ p})
      \right\}                              \\[1em]
      \left\{
      \exists w\ldotp \mathit{addr} \mapsto w \astm (\mathit{addr} \mapsto w \wand \mathit{PMP\_addr\_access\ entries\ p})
      \right\}
       & \,\mathit{return\_PMP\_ptsto}\,  &
      \left\{
      \mathit{PMP\_addr\_access\ entries\ p}
      \right\}
    \end{array}
  \end{math}
  \caption{Contracts for lemmas used in RISC-V PMP case study.}
  \label{fig:riscvlemmas}
\end{figure*}

To verify the memory integrity property for RISC-V with PMP, we use some interesting lemmas shown in \figref{fig:riscvlemmas}.
The first lemma $\mathit{open\_PMP\_entries}$ and a dual lemma called $\mathit{close\_PMP\_entries}$, open resp.\ close the $\mathit{PMP\_entries}$ predicate, to allow direct access to the PMP CSRs in parts of the ISA semantics that access them, particularly the PMP check algorithm.
We use the same scheme for reasoning about GPRs, \ie we pack them in a predicate and open and close it when appropriate.
The two lemmas needed for interacting with memory are $\mathit{extract\_PMP\_ptsto}$ and $\mathit{return\_PMP\_ptsto}$.
$\mathit{extract\_PMP\_ptsto}$ trades ownership of PMP-authorized memory, given by $\mathit{PMP\_addr\_access}$ for a points-to predicate for an authorized, in-range address $a$ and a magic wand that allows to recover \linebreak $\mathit{PMP\_addr\_access}$ using $\mathit{return\_PMP\_ptsto}$ if we return the points-to predicate.
All these lemmas are proven correct in the \Iris{} model and are explicitly invoked in its function definitions using ghost statements that aid the semi-automatic verification of the contracts of these functions by \Katamaran{}.

These lemma invocations suffice to let Katamaran verify most of the contracts in the codebase.
As for \MinimalCaps{}, we only need to prove the contract for the \textit{fdeCycle()} and the lemmas used in the case study manually.

\subsection{Evaluation}
\label{sec:evaluation}

In this section we evaluate our semi-automatic approach to universal contract verification.
Our aims for the approach are that universal contracts should be agnostic of software abstractions and verified against the operational semantics of ISAs.
Furthermore, we want to minimize the effort to re-verify a universal contract for a modified ISA.
We evaluate the proof effort required in our case study absolutely as well as relatively to Cerise~\citep{georges_cap_2021,georges2021efficient,van_strydonck_proving_2022}.

Cerise is close to our MinimalCaps case study because it establishes a very similar formulation of capability safety in an expressive, \Iris{}-based program logic.
However, they work for a simpler but otherwise similar capability machine ISA with a small-step operational semantics.
In addition to the universal contract which is formulated as a logical relation and associated "fundamental theorem", they also prove functional specifications for instructions which they use to verify example software.
The lines of code in both developments should be interpreted and compared with caution, because the approaches are quite different (e.g.\ different style of defining semantics) and because Cerise's functional specifications and verified examples do not have analogues in the MinimalCaps case study.
The Cerise proofs are large manual \Iris{} Proof Mode proofs, with limited automation.

\begin{table*}[t]
  \scriptsize
  \newcommand*\rot{\rotatebox{90}}
  \begin{center}
    \begin{tabular}{@{} l *{2}l | c | *{4}l | *{3}l | *{3}l | *{3}l | l || c @{} }
                              &                   &                  & Generatable         & \multicolumn{4}{c}{}   & \multicolumn{3}{|c}{Specs LoC} & \multicolumn{3}{|c|}{Proofs LoC} & \multicolumn{3}{c|}{Model and Solver} &                         & \multirow{2}{*}{\rot{Total LoC}\,\,\rot{(no \muSail{} or Oper. Sem.)}}                                                                                                                                                                                             \\
                              & \rot{\Sail{} LoC} & \rot{Oper. Sem.} & \rot{\muSail{} LoC} & \rot{\# \muSail{} Fns} & \rot{\# Foreign Fns}           & \rot{\# Lemmas}                  & \rot{\# Lemma Invoc.}                 & \rot{\muSail{} Fns LoC} & \rot{Foreign Fns LoC}                                                  & \rot{Lemma LoC} & \rot{\muSail{} Fns LoC} & \rot{Foreign Fns LoC} & \rot{Lemma LoC} & \rot{\Iris{} Model LoC} & \rot{\# Pure Pred.} & \rot{User Solver LoC} & \rot{Universal Contract LoC} & \\
      \textbf{\MinimalCaps{}} & 571               & -                & 1096                & 52                     & 3                              & 9                               & 39                                    & 397                     & 47                                                                     & 98             & 120                     & 65                    & 70              & 350                  & 3                   & 82                    & 172                            %& 2686
                              & 1508                                                                                                                                                                                                                                                                                                                                                                                                                                                                                           \\
      \textbf{RISC-V PMP}     & -                 & -                & 2025                & 66                     & 4                              & 8                                & 16                                    & 883                     & 61                                                                     & 89              & 226                     & 150                   & 390             & 167                  & 9                   & 178                   & 215                            %& 4188
                              & 2462                                                                                                                                                                                                                                                                                                                                                                                                                                                                                           \\
      \textbf{Cerise}         & -                 & 1190             & -                   & -                      & -                              & 142                              & -                                     & -                       & -                                                                      & 2318            & -                       & -                     & 2918            & 1351                 & -                   & -                     & -                              %& 7919
                              & 6729                                                                                                                                                                                                                                                                                                                                                                                                                                                                                           \\
    \end{tabular}
  \end{center}
  \caption{Detailed statistics for the \MinimalCaps{} and RISC-V PMP case studies and some comparative statistics (where relevant) with Cerise (the base version without uninitialized, local capabilities or I/O \citep{georges_cap_2021}), giving the lines of code (LoC) without comments for different parts of the case study as well as some numbers on how many \muSail{} functions, foreign functions, lemmas, lemma invocations and pure predicates each case study defined.
    There is no direct mapping of our approach to the approach taken for Cerise so the comparison is not entirely fair, for example, the \Iris{} model LoC for Cerise also contains code for verifying concrete code.
    We view the \Sail{} LoC as separate from our case studies and therefore do not include it the total at the right of the table.
  }
  \label{tab:stats}
\end{table*}

\tabref{tab:stats} presents statistics on our two case studies and some relevant statistics for Cerise \citep{georges_cap_2021,georges2021efficient,van_strydonck_proving_2022}.
We will first focus on the \MinimalCaps{} and RISC-V PMP rows of the table and discuss the comparison with Cerise at the end of this section.

The first column in the table shows the \Sail{} LoC for the \MinimalCaps{} case study.
For \MinimalCaps{} we started with our own \Sail{} specification and gradually extended it until it became a subset of CHERI-RISC-V.
We took the opposite direction for the RISC-V PMP case study, starting from the RISC-V \Sail{} specification and simplifying it during the translation step from \Sail{} to \muSail{} into a minimal subset with the PMP extension.
This means we do not have a simplified, minimal RISC-V PMP \Sail{} codebase and therefore do not report on the \Sail{} LoC for this case study.

The next part of the table is data about our case studies themselves.
Our case studies are based on \Sail{} codebases, which we currently manually translate into \muSail{} code, but we are confident that this translation can be automated.
The \muSail{} code is twice the size of the \Sail{} code, this due to some configuration that we need to provide for \Katamaran{} and the required derivation of typeclass instances.
Next, we present the number of \muSail{} functions, foreign functions, lemmas, and lemma invocations.
The lemmas aid \Katamaran{} in its verification endeavor and the invocations of these lemmas need to be manually added in the \muSail{} functions.
The contract proof LoC for the \muSail{} specification of our case study are indicative of how well \Katamaran{} was able to automate the boring parts.
For the \MinimalCaps{} case study, the majority of the 120 LoC for the \muSail{} specification proofs consists of tactic invocations to discharge trivial proof obligations.
This is similar for the RISC-V PMP case study, but \Katamaran{} left a few residual VCs that required manual discharging.
The proofs for the foreign functions and lemmas --- the interesting part of our case studies --- that reason about capability safety and memory interactions, require manual proof effort.
\Katamaran{} distinguishes between spatial and pure abstract predicates and provides hooks for a user-defined solver for pure predicates.
We specified a few pure predicates and report on the LoC for the solver.
Finally, we were able to validate our universal contracts in 172 LoC (\ie reasoning about the contract for the \emph{fdeCycle}) for \MinimalCaps{} and 215 LoC for RISC-V PMP.
We end \tabref{tab:stats} with the total LoC for the case studies, not including the \muSail{} specification (or operational semantics for Cerise), as the \muSail{} specification should in principle be generatable.

To further demonstrate the robustness of our approach to universal contracts, we have added an instruction to each case study that doesn't introduce any complexity regarding the proven universal contract, \ie we are adding a boring case to each case study.
We have chosen to duplicate the integer addition instruction for this purpose, which takes three registers, a destination register to write the result to and two source registers.
In RISC-V this means adding a new operation for the \emph{RTYPE} instructions, while for \MinimalCaps{} we define a completely new instruction.
The increase in the \muSail{} LoC specification is only two lines for the RISC-V case and 17 for \MinimalCaps{}.
No further changes are required for the RISC-V case, \ie we do not need to modify any proofs.
For \MinimalCaps{} we need to add a lemma invocation in the execute clause for the instruction and we need to specify a contract for the new instruction execution clause, which needs 3 LoC for the contract specification.
Furthermore, we add two lines of contract proof code to include this new instruction.
Due to the added lemma invocations, \Katamaran{} was able to verify the proof without further manual effort.
We conclude that adding a boring instruction (\ie an instruction that is not relevant for the universal contract) requires only minimal changes to both of our case studies.

The categories for which we provide statistics for \MinimalCaps{} have no direct mapping to Cerise, so for a meaningful discussion, we tried to gather statistics in a way that is maximally fair, but the reader should keep in mind that this comparison is not entirely fair.
For example, although we do not count Cerise proofs that are unrelated to the universal contract, the proofs do not separate lemmas and proofs that are intended for verifying concrete code from those that are used to prove the universal contract.
The operational semantics for the capability machine of Cerise is comparable to our \muSail{} specification.
More interesting are the statistics for lemmas and the lines of code for the specification and proofs of these lemmas, where Cerise requires significantly more proof effort.
For \MinimalCaps{}, we use \Katamaran{} to automate uninteresting parts, leaving us with a smaller amount of proof code for lemmas that are directly related to capability safety.
The \Iris{} model is already partially instantiated in the \Katamaran{} codebase, making the \MinimalCaps{} LoC for this part smaller than that of Cerise.

\section{Applying the universal contract: femtokernel verification}
\label{sec:femtokernel}
Thus far, we have focused on the verification of the security guarantees of our universal contracts.
In this section, we demonstrate that our universal contracts are strong enough to support the verification of properties of programs running on top of an ISA.
The \MinimalCaps{} UC is close to the Cerise model, for which this has arguably already been demonstrated \citep{van_strydonck_proving_2022,georges_cap_2021}.
Therefore, we focus on our RISC-V case, where, to the best of our knowledge, such a verification using universal contracts has not yet been demonstrated.

To illustrate the technique, consider the minimal \emph{femtokernel} in \figref{fig:femtokernel}, which configures the PMP extension to protect itself, including its interrupt handler~($\obj{\mathrm{ih}}$) and a private data field~($\obj{\mathrm{data}}$), from adversarial user mode code~($\obj{\mathrm{adv}}$).
More specifically, the femtokernel configures the PMP address registers to create the memory regions $[\obj{0},\obj{\mathrm{adv}})$ and $[\obj{\mathrm{adv}},\obj{\mathit{max}})$ (lines 1--4), where the $\obj{\mathit{max}}$ variable refers to the maximum size of memory available on the machine, then revokes all permissions for user mode from the first region and grants read, write and execute permissions to user mode to the second region (lines 5--6).
Both entries are unlocked so machine mode code can also access the first region.
The kernel then installs its handler (lines 7--8) and jumps to the adversary in user mode (lines 9--12) by loading the adversary address into the $\mathit{mepc}$ register (lines 9--10), clearing the $\mathit{mstatus}$ register (line 11), \ie setting the \emph{MPP} field to user mode, and performing the jump (line 12).
The handler will read the private data field into the $\mathit{ra}$ register before returning, but leaves the value in memory unchanged.

The integrity property we wish to verify is that the private $\obj{\mathrm{data}}$ field, which is initialized with value $42$, will always contain the value $42$ --- that is, code running in user mode cannot modify (or even directly read) the internal state of the kernel.
Our \Iris{} model supports a general notion of invariants, so we can register this property as an invariant (and in fact, we do the same for the memory storing the femtokernel instructions):
\begin{equation*}
  \mathit{inv}_{\mathit{femto}} = \knowInv{}{\mathit{data} \mapsto_{\mathrm{word}} 42} \qquad \text{ for data the label from Fig.~\ref{fig:femtokernel}.}
\end{equation*}
Ownership of the remaining memory is sufficient to establish PMP\_addr\_access\ entries\ \textrm{User} for the PMP configuration (\emph{entries}) set by the femtokernel initialization code.
As a result, we can invoke the universal contract to establish safety of the jumps to \emph{unknown user-mode code} at the end of the initialization and exception handler.
It then remains to prove contracts for the kernel and interrupt handler, establishing that they jump to user-mode code in a correct state (\ie in user-mode, with the intended PMP configuration) and use but don't break the registered invariants.

\begin{figure}[t]
  \begin{math}
    \begin{array}{r l l l}
      \linenumber{1:}  & \obj{\textrm{kernel:}} & \instr{la}    & \mathrm{ra}, \,\obj{\mathrm{adv}}             \\
      \linenumber{2:}  &                        & \instr{csrrw} & \mathrm{x0}, \,\mathrm{pmpaddr0}, \mathrm{ra} \\
      \linenumber{3:}  &                        & \instr{li}    & \mathrm{ra}, \,\obj{\mathit{max}}             \\
      \linenumber{4:}  &                        & \instr{csrrw} & \mathrm{x0}, \,\mathrm{pmpaddr1}, \mathrm{ra} \\
      \linenumber{5:}  &                        & \instr{li}    & \mathrm{ra}, \,\mathrm{0xf00}                 \\
      \linenumber{6:}  &                        & \instr{csrrw} & \mathrm{x0}, \,\mathrm{pmp0cfg}, \mathrm{ra}  \\
      \linenumber{7:}  &                        & \instr{la}    & \mathrm{ra}, \,\obj{\mathrm{ih}}              \\
      \linenumber{8:}  &                        & \instr{csrrw} & \mathrm{x0}, \,\mathrm{mtvec}, \mathrm{ra}    \\
      \linenumber{9:}  &                        & \instr{la}    & \mathrm{ra}, \,\obj{\mathrm{adv}}             \\
      \linenumber{10:} &                        & \instr{csrrw} & \mathrm{x0}, \,\mathrm{mepc}, \mathrm{ra}     \\
      \linenumber{11:} &                        & \instr{csrrw} & \mathrm{x0}, \,\mathrm{mstatus}, \mathrm{x0}  \\
      \linenumber{12:} &                        & \instr{mret}                                                  \\
      \linenumber{13:} & \obj{\textrm{ih:}}     & \instr{auipc} & \mathrm{ra}, \, 0                             \\
      \linenumber{14:} &                        & \instr{lw}    & \mathrm{ra}, 12(\mathrm{ra})                  \\
      \linenumber{15:} &                        & \instr{mret}  &                                               \\
      \linenumber{16:} & \obj{\textrm{data:}}   & 42            &                                               \\
      \linenumber{17:} & \obj{\textrm{adv:}}    & \ldots        &
    \end{array}
  \end{math}
  \caption{The femtokernel sets up the PMP entries to protect itself, the interrupt handler and its internal state.}
  \label{fig:femtokernel}
\end{figure}

Inspired by Islaris~\citep{islaris}, we can largely automate the remaining verification, by reusing existing components and proofs of \Katamaran{} to derive a sound verifier for known assembly code.
Essentially, the idea is that a contract for $\{\emph{Pre}\}~ \emph{B}~ \{\emph{Post}\}$ for a basic
block of assembly code can also be regarded as a contract for the ISA semantics, under the
assumption that it is looking at basic block $B$.
Essentially, to verify a contract for a basic block of assembly instructions $\{ \mathit{Pre}_0 \}~ \overline{\mathit{instr}}_{0..n}~ \{ \mathit{Post}_n\}$, we iteratively verify \textit{fdeStep()} contracts that look roughly like this:
\begin{multline*}
  \{ \mathit{Pre}_i \ast \mathrm{pc} \mapsto a \ast a \mapsto c \ast decode(c) = \mathit{instr}_i \}~ \mathit{fdeStep()}\\~ \{ \mathit{Post}_i \ast \mathrm{pc} \mapsto a \ast a \mapsto c \}
\end{multline*}
We are thus able to verify the contracts for the basic blocks of the femtokernel, \ie the initialization and handler code, leaving only some manual proofs to register invariants for code, data, and adversarial memory and invoke the universal contract.

Taken together, our femtokernel case study demonstrates that our UC for RISC-V can be directly applied for verifying security properties of trusted code relying on PMP to interact with untrusted code.
All parts of the verification are fully verified in \Coq{}, yielding a rigorous proof about ISA execution, directly in terms of \muSail{}'s operational semantics, which we list for reference in Appendix~\ref{sec:end-end-security}.

\section{Related Work}
\label{sec:related-work}
Universal contracts have been used to capture security properties of capability-based high-level languages
\citep{devriese2016reasoning,swasey2017robust} and capability-based ISAs \citep{skorstengaard2018reasoning,georges2021efficient,georges_cap_2021,georges_cap_2021,van2019linear}.
Our formalization of capability safety for \MinimalCaps{} is close to the one in Cerise \citep{georges2021efficient,georges_cap_2021,georges_cap_2021}.
Some versions of Cerise support additional features like local or uninitialized capabilities.
They use a verification approach that requires significant effort to prove that the universal contracts hold, in contrast to our semi-automatic verification approach enabled by \Katamaran{}.

\Citet{nienhuis2020rigorous} prove \emph{reachable capability monotonicy} up to security domain transitions and intra-domain memory invariant properties for the entire CHERI-MIPS ISA, based on the L3 specification instead of the \Sail{} specification, where their security property is based on the ISA specification and does not take a hardware implementation or software running on the ISA into account.
There are some differences between their work and ours: first, we have demonstrated that the \emph{capability safety} security property we formulate as a universal contract can be used in the verification of programs to be executed on the ISA.
It has not been shown yet that capability monotonicity up to security domain transitions is a strong enough property to perform such full-system proofs.
Second, the approach taken differs from ours in that they express their security
property in the meta-logic directly and automate the \emph{boring} parts of the proof away with standard automation, like tactics and auto-generated proof scripts, whereas we use an embedded separation logic and provide our semi-automatic logic verifier, \Katamaran{}, based on symbolic execution.
Our more abstract description of the security property should be more future proof against ISA modifications and extensions.
For example, in our universal contracts it does not matter whether a capability machine has a merged or split register file for capability registers, whereas \citeauthor{nienhuis2020rigorous} mention that such a change required refactoring of the properties and proofs in their approach.
Finally, we demonstrate the generality of our universal contracts approach by verifying security properties of non-capability machines.

Similar work to that of \citeauthor{nienhuis2020rigorous} is done by \citet{bauereiss2022morello} on a full-scale industry architecture, Morello, implementing the CHERI extension.
To reason about the ISA, a translation from the Arm ASL specification to \Sail{} occurs first, and from the \Sail{} specification it is possible to generate code for proof assistants such as Isabelle and \Coq{}.
To verify the reachable capability monotonicity property (up to domain
transitions), the authors define four properties of arbitrary CHERI instruction
execution and use that to verify a concrete implementation, i.e. Morello.
In comparison to \citeauthor{nienhuis2020rigorous}, this more abstract definition of the security property allows more automation and thereby reach the scale of Morello.
\Citeauthor{bauereiss2022morello} mention that proving stronger properties, such as capability safety, requires proof techniques that do not scale up to full-scale industry architectures.
Part of the reason is that current automation techniques for separation logics in a foundational setting~\cite{bedrock,chargueraud2020,diaframe,mosel,refinedc,vstfloyd} are still insufficient.
This is the issue that we are addressing with our proposed universal contract methodology and \Katamaran{} to semi-automatically verify universal contracts.

\citet{gao2021end} formally verify the correct execution of CHERI-instructions and liveness
properties for CHERI-Flute \citeyearpar{cheriflute}: a concrete implementation of CHERI-RISC-V.
Their results apply only to this specific ISA implementation, making their work very different from ours.
Similarly, \citet{cheang2020verifying} prove functional correctness of RISC-V PMP for the Rocket Chip implementation, as a step towards verifying the Keystone \citep{lee2020keystone} framework.

\Citeauthor{dam2013formal} and \citeauthor{khakpour2013machine} use a basic Hoare logic to formulate ISA security primitives for Armv7.
This suffices for their security primitive, where the CPU transitions only from untrusted code to trusted code on interrupt or syscall, making it easy to define the scenarios for which trusted code needs to ensure secure behavior.
Capability safety exemplifies a more complicated security property because arbitrarily many security boundaries can be crossed: any enter capability that is directly or indirectly accessible from current architectural registers.
Because of this, phrasing such guarantees requires a sufficiently expressive logic.
Other security primitives can be expected to similarly require or benefit from advanced logic features like ghost state, invariant, ownerhsip etc.

\citet{guarnieri2021contracts} propose hardware/software contracts to formalize security guarantees in a minimal ISA setting that takes side-channel attacks into account.
A similar approach is taken by \citet{ge2018no}, who propose adding guarantees about side-channel
leakage to the ISA in an \emph{augmented ISA (aISA)}.
Both proposals address a different problem than we do: while we leave confidentiality guarantees, microarchitectural aspects and side-channel leakage out of scope, they do the same with security boundaries, architectural security primitives and direct-channel protections.
In that sense, they are formalizing a different aspect of ISA security guarantees, which should ultimately be combined with direct-channel guarantees like ours to obtain a complete ISA security specification.
In future work, we intend to add support for confidentiality guarantees and side channels.

The conventional way to reason about separation logic in proof assistants is to use a shallow embedding of propositions and provide meta-programming facilities, like tactics or plugins, which can be used to implement proof steps that \emph{interactive forward symbolic execution}~\cite{vstfloyd} of program fragments at the meta-level~\cite{bedrock,chargueraud2020,diaframe,mosel,refinedc}.
In the background a proof term is constructed which has to be checked by the system for each run.
In contrast, \Katamaran{} is not interactive and uses a deep embedding of propositions and is implemented at the object-level language of \Coq{} called Gallina and we verified once and for all that each run of the symbolic executor is sound.
As a consequence, the usage and implementation of \Katamaran{} is closer to standalone verifiers and frameworks~\cite{dafny,schwerhoff,verifast,viper,why3}.
A downside of the approach is that existing meta-level machinery cannot be used during symbolic execution for automation including simple symbolic evaluation, which instead have to be specifically implemented.
The deep embedding also incurs an increased complexity in terms of explicitly dealing with logic variable allocation.
VeriSmall~\cite{verismall} and Mechanized Featherweight VeriFast~\cite{fvf} are to the best of our knowledge the only other mechanized symbolic executors for separation logic based on deep embeddings.
However, both largely serve as a proof of concept mechanizations for practical systems~\cite{smallfoot,verifast} while \Katamaran{} aspires to be a practical mechanized implementation in itself.

\section{Conclusion}
\label{sec:conclusion}
Our work shows that universal contracts, together with \Katamaran{}, form a compelling, tool-supported method for formalizing, verifying and applying the security guarantees of ISAs.
We have demonstrated that the approach applies to ISAs with very different security primitives (capabilities versus PMP+exceptions), to industrially relevant ISAs like RISC-V PMP (with only minor simplifications remaining) and that the approach balances the requirements of ISA designers and authors of security-critical software.
We have demonstrated how we can check the consistency of universal contracts with functional ISA semantics and how they can be applied to reason about security-critical software.
Finally, we have shown that \Katamaran{} is able to effectively semi-automate the verification of universal contracts as well as the verification of security-critical software, offering convenient ways for users to inject manually proven lemmas in an automatic verification.

In other words, our current results already demonstrate the viability and generality of the approach to verify ISA security properties using universal contracts and \Katamaran{}.
In the future, we intend to scale up our approach by applying it to larger ISAs, supporting complex semantic features (such as asynchronous interrupts, concurrency etc) and other security properties (e.g.\ confidentiality).
We also intend to automate the translation from \Sail{} to \muSail{} and improve \Katamaran{} automation further.

% -------------------------------------------------------------------------------
\section*{Acknowledgments}
% -------------------------------------------------------------------------------

This research was partially funded by the Research Fund KU Leuven, by the Flemish Research Programme Cybersecurity and by a European Research Council (ERC) Starting Grant (UniversalContracts; 101040088), funded by the European Union.
Views and opinions expressed are, however, those of the authors only and do not necessarily reflect those of the European Union or the European Research Council.
We thank Thomas Van Strydonck for his help gathering statistics and Denis Carnier for proof-reading.

% The USENIX latex style is old and very tired, which is why
% there's no \textbackslash{}acks command for you to use when
% acknowledging. Sorry.

% -------------------------------------------------------------------------------
\section*{Availability}
% -------------------------------------------------------------------------------

The source code of the \Coq{} development for this paper is available at \href{https://github.com/katamaran-project/katamaran/releases/tag/ccs23}{\color{blue}{https://github.com/katamaran-project/katamaran/releases/tag/ccs23}}.

\bibliographystyle{plainnat}
\bibliography{main}

\appendix

\section{End-to-end security statement about Femtokernel}
\label{sec:end-end-security}

\begin{align*}
  \mathrm{Lemma}~& \mathit{femtokernel\_endToEnd} :\\
  &\mathit{mem\_has\_instrs}~ \mu~ 0~ \mathit{femtokernel\_init} \rightarrow\\
  &\mathit{mem\_has\_instrs}~ \mu~ 72~ \mathit{femtokernel\_handler} \rightarrow\\
  &\mathit{mem\_has\_word}~ \mu~ 84~ 42 \rightarrow\\
  &\mathit{read\_register}~ \gamma~ \mathit{cur\_privilege} = \mathit{Machine} \rightarrow\\
  &\mathit{read\_register}~ \gamma~ \mathit{pmp0cfg} = \mathit{femtokernel\_default\_pmpcfg} \rightarrow\\
  &\mathit{read\_register}~ \gamma~ \mathit{pmpaddr}_0 = 0 \rightarrow\\
  &\mathit{read\_register}~ \gamma~ \mathit{pmp1cfg} = \mathit{femtokernel\_default\_pmpcfg} \rightarrow\\
  &\mathit{read\_register}~ \gamma~ \mathit{pmpaddr}_1 = 0 \rightarrow\\
  &\mathit{read\_register}~ \gamma~ \mathit{pc} = 0 \rightarrow\\
  &\langle \gamma, \mu, \delta, \mathit{fdeCycle}() \rangle \longrightarrow^* \langle \gamma', \mu', \delta', s' \rangle \rightarrow\\
  &\mu'~ 84 = 42.
\end{align*}

\end{document}